\documentclass[a4paper,11pt]{article}
\usepackage{amsfonts}
\usepackage{amsmath}
\usepackage{amssymb}
\usepackage{graphicx}
\usepackage{array}
\usepackage{booktabs}
\usepackage{multirow}
\usepackage{makecell}
\usepackage{float}
\pdfoutput=1 

\usepackage{mymacros}
\usepackage{chngcntr}
\usepackage{verbatim}
\usepackage{amsthm}
\usepackage{graphics}
\usepackage{xcolor}
\usepackage{mathrsfs}
\usepackage{bbold}
\usepackage{cases}
\usepackage{epsfig}
\usepackage{epstopdf}
\usepackage{hyperref}
\usepackage{amsfonts}
\usepackage{hyperref}
\usepackage{jheppub} 

\usepackage[T1]{fontenc} 

\hypersetup{
	colorlinks=true,
	linkcolor=blue,
	filecolor=blue,
	urlcolor=blue,
	citecolor=blue,
}

\title{\boldmath Testing the chaos bound in the spinor field of Einstein-Euler-Heisenberg-Anti-de Sitter spacetime}

\author{Xiaowei Li $^{a}$,}
\author{Bingbing Chen $^{a}$ and}
\author{Guoping Li $^{b}$$\footnote{\texttt{Corresponding author:gpliphys@yeah.net}}$}

\affiliation{$^{a}$ School of Mathematics, Physics and Statistics, Sichuan Minzu College, Ganzi 626100, People's Republic of China\\
$^{b}$ Physics and Astronomy College, China West Normal University, Nanchong 637000, People's Republic of China}

\abstract{Violations of the chaos bound have been observed in scalar fields. In this work, we investigate the Lyapunov exponents of chaotic particles in the spinor field of an Einstein-Euler-Heisenberg-Anti-de Sitter spacetime, and test the validity of the chaos bound in this field. The influences of the black hole charge, the Euler-Heisenberg constant, cosmological constant, particle charge and total angular momentum on the exponents are analyzed. With other parameters fixed, chaos bound violations occur only within specific ranges of black hole charge, particle spin, or total angular momentum-unlike in Reissner-Nordstr\"om spacetimes, where violations intensify with larger parameter values. Notably, anti-alignment of particle spin with the $z$-axis can trigger violations even for small cosmological constants, while no violations arise for the spin aligned with the $z$-axis regardless of the cosmological constant. Our results show that the cosmological constant drastically reshapes chaos bound violation conditions compared to Reissner-Nordstr\"om spacetimes, highlighting the spin's pivotal role in chaos onset. }

\keywords{Spinning particles, chaos bound, cosmological constant, EEH-AdS black hole}

\begin{document}
	\maketitle
	\flushbottom

\section{Introduction}\label{sec1}

In 2016, Maldacena, Shenker, and Stanford proposed a conjecture: In thermal quantum systems with a large number of degrees of freedom, the Lyapunov exponent(LE) is bounded by a temperature-dependent limit~\cite{Maldacena:2015waa},

\begin{equation}
\lambda \leq \frac{2\pi T}{\hbar}, \label{eq:bound}
\end{equation}

\noindent where $\lambda$ denotes the exponent and $T$ is the system temperature. This bound(chaos bound) was established via the AdS/CFT correspondence~\cite{Maldacena:1997re} through thought experiments involving shock waves near the black hole horizon~\cite{Shenker:2013pqa}. Since then, this conjecture has been extensively investigated. In Einstein gravity, the bound is saturated for black hole systems~\cite{Shenker:2014ywa}, and saturation also occurs in the Sachdev-Ye-Kitaev (SYK) model~\cite{SYK1,SYK2,SYK3,SYK4}. Studies of classical chaos for particles outside a black hole indicate that the exponent obeys $\lambda \leq \kappa$, where $\kappa$ is the surface gravity, independent of the external forces and the particle mass~\cite{Hashimoto}.

Despite widespread support for this bound, exceptional cases of its violation have been identified. In the radial direction, the authors analyzed the static equilibrium conditions of charged particles, solved for their effective potentials, and expanded the exponents describing unstable orbits near the horizons, thereby examining the relationship between these exponents and the surface gravities of the black holes. It was found that while the bound holds in Reissner-O ordstr\"om (RN) and Reissner-Nordstr\"om-Anti-de Sitter (RN-AdS) spacetimes, violations arise in certain alternative spacetimes \cite{ZLL}. When the contribution of a particle's orbital angular momentum is taken into account, violations are also observed in the RN and RN-AdS spacetimes \cite{LG1}. Furthermore, violations have been observed in other spherically symmetric spacetimes \cite{LG2,LG3,LG4,GC1,S1,GC2,S2,S3}, as well as in axisymmetric spacetimes including Kerr-Newman \cite{KG1}, Kerr-Newman-dS \cite{KG2}, Kerr-Newman-AdS \cite{KG3,KG6}, and Kerr-Sen-AdS black holes \cite{KG5}. Other instances of such violations are found in \cite{SSS1,SSS2,SSS3,SSS4,KG4,KG7}.

The aforementioned violations were first identified in scalar field systems. Recently, Yang et al. examined this bound in the spinor field within RN spacetime \cite{Yang2026}. They found that under certain conditions, the classical chaos exponents of both neutral and charged spinning particles can violate the bound. Notably, the black hole charge emerges as a critical parameter: violations occur exclusively when the charge is sufficiently large, with no violations observed for the small charge values. However, the RN spacetime considered in \cite{Yang2026} is asymptotically flat. In more physically motivated cosmological scenarios, a non-vanishing cosmological constant modifies the asymptotic structure of spacetime and may thus influence the onset of chaos. Additionally, quantum electrodynamics (QED) corrections-such as the Euler-Heisenberg (EH) term-alter the electromagnetic field configuration around charged black holes. It is therefore imperative to investigate how these factors impact the violation of the bound by spinning particles.

In this paper, we investigate the validity of the chaos bound for spinor fields in Einstein-Euler-Heisenberg-Anti-de Sitter (EEH-AdS) spacetime by numerically computing LEs of chaotic spinning particles orbiting EEH-AdS black holes. The EEH-AdS black hole is a charged black hole solution derived by coupling QED vacuum polarization effects to gravitational theory within the framework of nonlinear electrodynamics (NLED) \cite{MB}. Its defining feature is the incorporation of quantum correction terms from the EH effective field theory, which extends the classical RN-AdS black hole. These corrections arise from the nonlinear self-interaction of the electromagnetic field, induced by electron-positron pair fluctuations in the quantum vacuum. Due to the presence of the EH constant, this black hole exhibits stronger gravitational attraction at short distances compared to its RN-AdS counterpart, rendering the system's behavior more similar to that of a Schwarzschild black hole than the RN-AdS black hole \cite{MB2}. Previous studies have shown that such black holes can undergo two distinct types of phase transitions: one driven by the nonlinear electromagnetic field, and the other corresponding to a gravitational phase transition \cite{LGL1,LGL2}.

The remainder of this paper is organized as follows. In Section~2, we derive the equations of motion for spinning particles in the EEH-AdS spacetime. In Section~3, we discuss the influence of the black hole charge, particle charge, total angular momentum, and spin on the LEs, and examine the chaos bound in the spinor field within the EEH-AdS spacetime. Section~4 presents our conclusions.

\section{Derivation of LEs in EEH-AdS spacetime}\label{sec2}

The four-dimensional action describing general relativity with cosmological constant coupled to the EH NLED reads \cite{JFP1,JFP2}

\begin{eqnarray}
S=\frac{1}{4\pi}\int_{M^4}d^4x\sqrt{-g}\left[\frac{1}{4}(R-2\Lambda) -\mathcal{L}(F,G) \right].
\label{eq2.1}
\end{eqnarray}

\noindent In the above action, $R$ is the Ricci scalar and $\Lambda$ denotes the cosmological constant, which can take either positive or negative values. In this paper, we focus on the AdS spacetime, where the cosmological constant is negative. The Lagrangian density $\mathcal{L}(F,G)$ for the EH NLED is given by \cite{JFP3}

\begin{eqnarray}
\mathcal{L}(F,G) = -F+\frac{1}{2}aF^2 +\frac{7}{8}aG^2,
\label{eq2.2}
\end{eqnarray}

\noindent where $F = \frac{1}{4}F_{\mu\nu}F^{\mu\nu}$ and $G = \frac{1}{4}F_{\mu\nu}{\star F}^{\mu\nu}$ are the electromagnetic invariants constructed from the field strength tensor $F_{\mu\nu}$. The parameter $a = \frac{8\alpha^2}{45m^4}$ denotes the EH coupling constant, which governs the strength of the NLED contribution; here, $\alpha$ is the fine-structure constant and $m$ is the electron mass.

The EEH-AdS black hole is a static, spherically symmetric solution derived from the above action, with the metric given by \cite{MB}.

\begin{eqnarray}
ds^2 &=& -f(r)dt^2 + \frac{1}{f(r)}dr^2 + r^2(d\theta^2+\sin^2\theta d\phi^2),\nonumber\\
f(r) &=& 1-\frac{2M}{r}+\frac{Q^{2}}{r^{2}}-\frac{\Lambda r^{2}}{3}-\frac{aQ^{4}}{20r^{6}}
\label{eq2.3}
\end{eqnarray}

\noindent with an electromagnetic potential

\begin{eqnarray}
A_t=\frac{Q}{r}-\frac{aQ^{3}}{10r^{5}},
\label{eq2.4}
\end{eqnarray}

\noindent where $M$ and $Q$ denote the mass and charge of the black hole, respectively. For $\Lambda = 0$, the metric reduces to that of the charged EEH black hole derived by Amaro and Macias \cite{AM,YT1,YT2}. In the limit $a = 0$, the metric describes the RN-AdS black hole. When $a=0$ and $\Lambda = 0$, the metric is reduced to the RN metric; for $Q=0$, it further reduces to the Schwarzschild metric. Outside the event horizon, the term $\frac{Q^2}{r^2}$ dominates over $\frac{a Q^4}{20r^6}$, and the parameter $a$ is constrained to the range $0 \leq a \leq \frac{32Q^2}{7}$ o ensure the solution remains within the regime of linear electrodynamics \cite{MB}. The event horizon $r_+$ is defined by the condition $f(r) = 0$, with the corresponding surface gravity given by

\begin{eqnarray}
\kappa=\frac{1}{2r_+}\left(1 - \frac{Q^2}{r_+^2}-\Lambda r_+^2+\frac{aQ^4}{4r_+^6}\right).
\label{eq2.5}
\end{eqnarray}

In this work, we consider a spinning particle of mass $m$ and charge $q$ in this spacetime, which obeys the Mathisson-Papapetrou-Dixon (MPD) equations \cite{HH1}

\begin{align}
\frac{Dp^\mu}{D\tau} &= -\frac12 R^\mu_{\nu\alpha\beta} u^\nu S^{\alpha\beta} - m q F^\mu_{\ \nu} u^\nu, \label{eq:mpd1}\\
\frac{DS^{\mu\nu}}{D\tau} &= p^\mu u^\nu - u^\mu p^\nu, \label{eq:mpd2}
\end{align}

\noindent where \(\frac{D}{D\tau}\) denotes the covariant derivative along the trajectory of the particle, and \(\tau\) is the affine parameter. The four-momentum and four-velocity vectors of the particle are denoted by \(p^\mu\) and \(u^\mu = \frac{dx^\mu}{d\tau}\), respectively. The Riemann tensor is given by \(R^\mu_{\ \nu\alpha\beta}\), and \(S^{\mu\nu}\) represents the antisymmetric spin tensor. The electromagnetic field strength tensor of the background spacetime is given by \(F_{\mu\nu} = A_{\mu,\nu} - A_{\nu,\mu}\). To establish a well-defined relation between the four-velocity and the four-momentum of the particle, it is necessary to impose a spin supplementary condition. In this work, we adopt the Tulczyjew-Dixon spin supplementary condition (TDSSC) \cite{HH2,HH3},

\begin{equation}
S^{\mu \nu}p_{\nu} = 0.\label{eq2.6}
\end{equation}

\noindent The spin tensor \( S^{\mu\nu} \) and the four-momentum \( p^{\mu} \) are characterized by the conserved mass \( m \) and spin magnitude \( \tilde{S} \), respectively, and obey the following relation,

\begin{eqnarray}
m^2 = (p^t)^2 f - \dfrac{(p^r)^2}{f} - r^2 (p^\phi)^2, \quad
\tilde{S}^2 = \dfrac{r^2 (S^{r\phi})^2}{f}-(S^{tr})^2 - f r^2 (S^{t\phi})^2 .\label{eq2.7}
\end{eqnarray}

\noindent We adopt the method developed in \cite{NZ} to solve the particle's equations of motion. First, expanding the TDSSC condition in the EEH-AdS spacetime, we obtain

\begin{equation}
\begin{aligned}
\dfrac{p^r S^{tr}}{f} + r^2 p^\phi S^{t\phi} = 0, \label{eq2.9}
\end{aligned}
\end{equation}
\begin{equation}
\begin{aligned}
r^2 p^\phi S^{r\phi} + f p^t S^{tr} = 0,\label{eq2.10}
\end{aligned}
\end{equation}
\begin{equation}
\begin{aligned}
\dfrac{1}{f} p^r S^{r\phi} - f p^t S^{t\phi} = 0. \label{eq2.11}
\end{aligned}
\end{equation}

\noindent The momentum equations \eqref{eq:mpd1} can be explicitly formulated as follows,

\begin{equation}
\begin{aligned}
\dot{p}^t +\frac{p^r f'}{2f} +\frac{\dot{r}p^t f'}{2f} -\frac{r\dot{\phi}S^{t\phi}f'}{2} -\frac{\dot{r}S^{tr}f'}{2} +\frac{m\dot{r}qA_{t,r}}{f} &= 0, \label{eq2.12}
\end{aligned}
\end{equation}
\begin{equation}
\begin{aligned}
\dot{p}^r +\frac{p^t f f'}{2} -\dot{\phi} p^\phi f' r -\frac{r\dot{\phi}S^{r\phi}f'}{2} -\frac{S^{tr}f f'}{2} -\frac{\dot{r}p^r f'}{2f} +\frac{\dot{r}qA_{t,r}}{f} +mqA_{t,r}f &= 0,\label{eq2.13}
\end{aligned}
\end{equation}
\begin{equation}
\begin{aligned}
\dot{p}^\phi +\frac{\dot{\phi}p^r}{r} +\frac{\dot{r}p^\phi}{r} -\frac{\dot{r}S^{r\phi}f'}{2fr} -\frac{S^{t\phi}f f'}{2r} &= 0, \label{eq2.14}
\end{aligned}
\end{equation}

\noindent In the above equations, the dot denotes differentiation with respect to coordinate time, while the prime denotes differentiation with respect to the radial coordinate. The spin equation is derived from Eq. \eqref{eq:mpd2} as

\begin{equation}
\begin{aligned}
\dot{S}^{tr} + p^r - \dot{r} p^t - f r \dot{\phi} S^{t\phi} &= 0,\label{eq2.15}
\end{aligned}
\end{equation}
\begin{equation}
\begin{aligned}
\dot{S}^{t\phi} - \dot{\phi} p^t + p^\phi + \frac{\dot{\phi} S^{tr} + \dot{r} S^{t\phi}}{r} + \frac{S^{r\phi} f' + \dot{r} S^{t\phi} f'}{2f} &= 0, \label{eq2.16}
\end{aligned}
\end{equation}
\begin{equation}
\begin{aligned}
\dot{S}^{r\phi} - \dot{\phi} p^r + \dot{r} p^\phi + \frac{\dot{r} S^{r\phi}}{r} + \frac{S^{t\phi} f f'}{2} - \frac{\dot{r} S^{r\phi} f'}{2f} &= 0.\label{eq2.17}
\end{aligned}
\end{equation}

\noindent To derive the explicit expression for the momentum, we invoke two conserved quantities: the particle's energy ($E$) and its total angular momentum ($L$) along the $z$-axis.

\begin{align}
E = f p^{t} - \frac12 f' S^{tr} + m q A_t,\quad
L = r^{2}p^{\phi} + rS^{r\phi}. \label{eq2.18}
\end{align}

\noindent Solving Eqs. \eqref{eq2.7}-\eqref{eq2.18} simultaneously, we obtain

\begin{align}
\dot r = \frac{p^{r}}{p^{t}}, \label{eq2.20}
\end{align}

\noindent where

\begin{align}
p^{t} &= -\frac{2m^{2} r (E - qA_t) \mp m\tilde S L f'}{(2m^{2} r - \tilde S^{2} f') f}, \quad p^{\phi} = -\frac{m^2L \mp m(E - qA_t)\tilde S}{\frac{1}{2}r \tilde S^{2} f' - m^2 r^{2}} \label{eq2.21}\\
p^{r} &= \pm \sqrt{ p_t^{2} - f\!\left(m^{2} + \frac{p_\phi^{2}}{r^{2}}\right) }. \label{eq2.22}
\end{align}

\noindent For the ``$\pm$'' symbol in Eq. \eqref{eq2.20}, the negative and positive signs correspond to directions opposite to and aligned with the $z$-axis, respectively. As for the ``$\mp$'' symbol in Eq. \eqref{eq2.21}, the positive and negative signs denote the outgoing and incoming directions of the particle, respectively.

The derivation of LEs for scalar and spinor particles is well-documented in  \cite{SSS4,Yang2026,HG1,HG2}. We employ its expression \cite{Yang2026,HG2},

\begin{equation}
\begin{aligned}
\lambda^2  = -\frac{V_{\text{eff}}''(r_0)}{m} = \frac{1}{2}\frac{d^2}{dr^2}\left(\frac{p^r}{p^t}\right)^2\bigg|_{r = r_0},\label{eq2.23}
\end{aligned}
\end{equation}

\noindent to numerically examine the relationship between its value and the surface gravity. In the above equation,

\begin{equation}
\begin{aligned}
V_{\text{eff}}(r_0)=-\frac{m}{2}\left(\frac{p^r}{p^t}\right)^2\bigg|_{r = r_0},\label{eq2.24}
\end{aligned}
\end{equation}

\noindent is an effective potential evaluated at an equilibrium position $r_0$.

\section{Examination of the chaos bound in EEH-AdS spacetime}\label{sec3}

The effective potential of particles encodes their dynamical behavior in strong gravitational fields and facilitates the classification of particle orbit types. We begin by analyzing the radial dependence of the effective potential.

\begin{figure}[h]
	\centering
	\includegraphics[width=10cm,height=7cm]{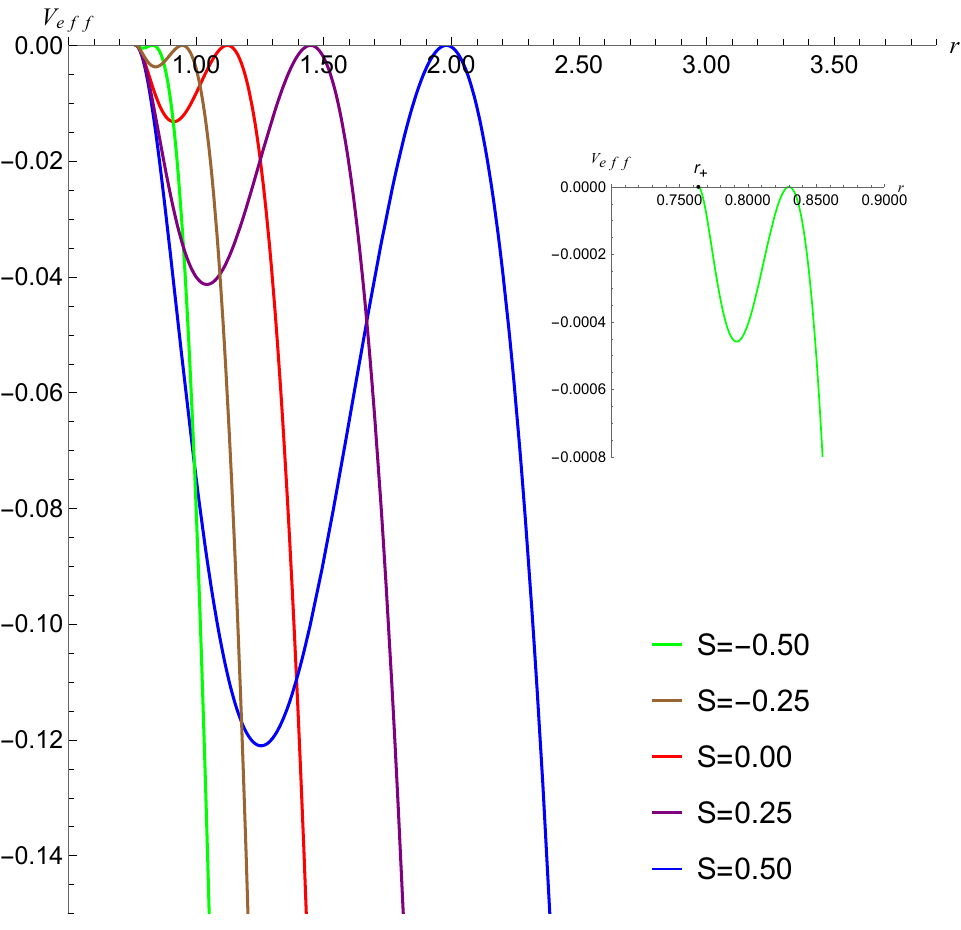}
	\caption{The effective potential of the particle varies with its radius. In our calculations, we set the parameters as follows: $Q=0.80$, $L=10.00$, $\Lambda=-3.00$, $q=-10.00$ and $a=0.60$.}
	\label{4f1}
\end{figure}

Figure \ref{4f1} plots the effective potential as a function of radius for particles with distinct spins. Inspection of the figure reveals that each effective potential curve exhibits two local maxima: one located at the event horizon, and the other outside. Here, we focus exclusively on the latter. When the spin direction is antiparallel to the angular momentum direction, the radial positions of the outer maxima (i.e., the equilibrium orbital radii) lie closer to the event horizon than when the two directions are parallel. For the non-spinning particle, the position of this maximum falls between these extremal cases. These equilibrium orbital radii and spin magnitudes directly influence the value of the LEs, thereby affecting the violation of the chaos bound.

\begin{figure}[h]
	\centering
	\includegraphics[width=10cm,height=7cm]{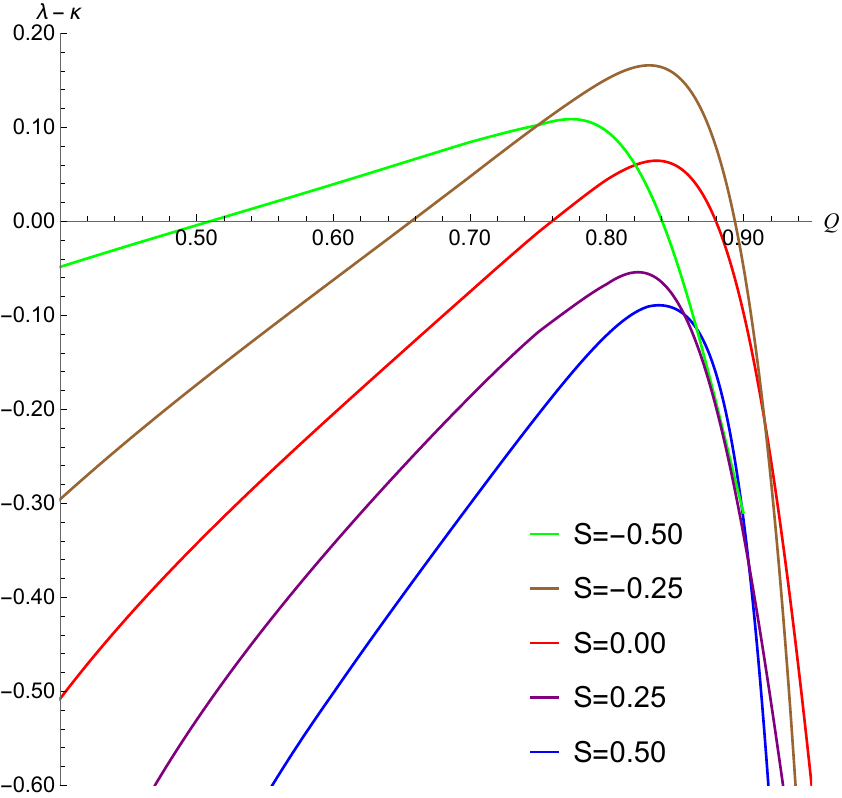}
	\caption{The relationship between the violation of the chaos bound and the black hole's charge. In the calculations, we set the parameters as follows: $L=10.00$, $\Lambda=-3.00$, $q=-10.00$ and $a=0.60$.}
	\label{4f2}
\end{figure}

Using Eqs. \eqref{eq2.3}-\eqref{eq2.5} and \eqref{eq2.21}-\eqref{eq2.23}, we numerically calculate the relationship between the LEs and surface gravity to determine whether the bound is violated. 
If $\lambda > \kappa$, the bound is violated. Otherwise, there is no violation.
Throughout all calculations, we set $M=m=1$ without loss of generality, and additionally impose the constraint that the particle's spin magnitude is less than the black hole mass \cite{RMW1,RMW2}. In our calculations, since $\tilde{S}$ denotes the spin magnitude of the particle, we redefine the spin as $S=\pm\tilde{S}$ to investigate the effect of spin orientation on the LEs. Here, the positive and negative signs correspond to spin directions aligned with and opposite to the $z$-axis, respectively.

We first elucidate the relationship between chaos bound violations and black hole charge in Figure \ref{4f2}. As is evident from the figure, for the particle with distinct spins, the differences between the exponent and surface gravity first increase and then decrease with increasing black hole charge, exhibiting clear maxima. Notably, when the particle's spin direction is parallel to its angular momentum, these maxima remain below the surface gravity, such that no violation of the bound occurs. Violations only emerge when the two directions are anti-parallel or the particle is spinless. Furthermore, the range of charge values over which violations occur for the anti-parallel spin orientations is broader than that for the spinless particle. Thus, the spin can either expand or narrow the charge range corresponding to bound violations. To streamline subsequent discussions of chaos bound violations, we fix the black hole charge at $0.80$ for all remaining calculations.

\begin{figure}[h]
	\centering
	\includegraphics[width=10cm,height=7cm]{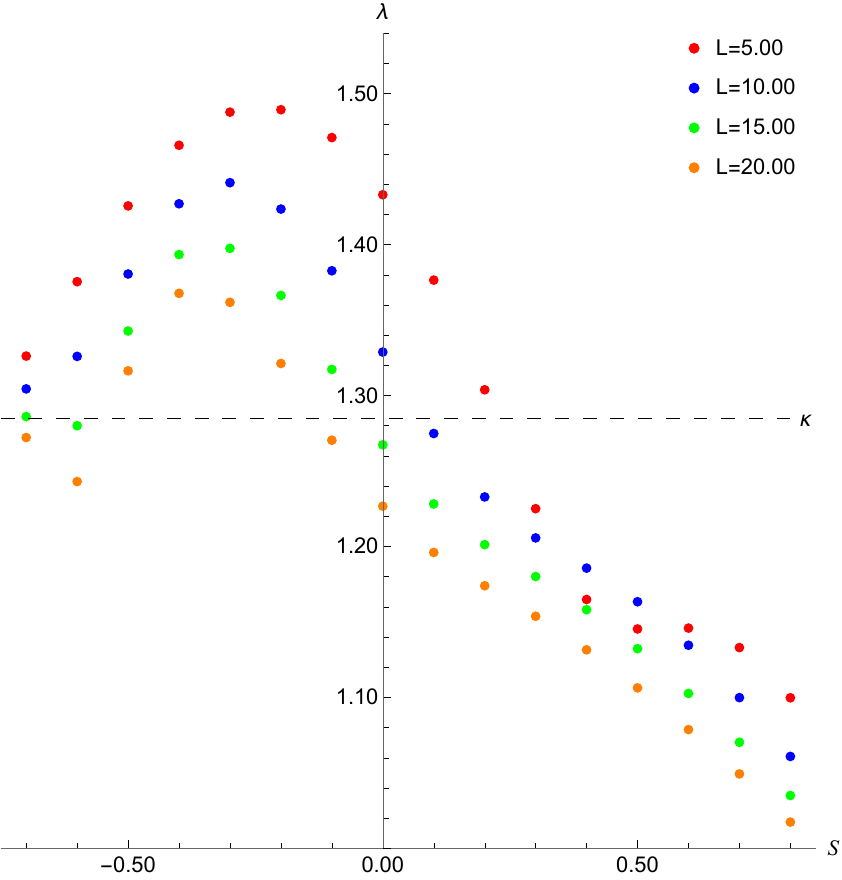}
	\caption{LEs vary with the spin of the particle with different values of total angular momentum.  In the calculations, we set the parameters as follows: $\Lambda=-3.00$, $q=-10.00$ and $a=0.60$.}
	\label{4f3}
\end{figure}

We present a detailed analysis of how the spin affects the exponents. As illustrated in Figure \ref{4f3}, when the particle's spin direction is parallel to its angular momentum, the exponents values decrease monotonically with increasing spin magnitude across all angular momentum cases considered. Specifically, for $L=15.00$ and $20.00$, the exponents values remains consistently below the surface gravity, such that no bound violation occurs. In contrast, for $L=5.00$ and $10.00$, the exponents exceed the surface gravity-resulting in a bound violation-only when the spin magnitude is below a critical threshold. Conversely, when the spin direction is anti-parallel to the angular momentum, the exponents first increase and then decrease with increasing the spin magnitude, with bound violations observed across all angular momentum values studied. Notably, the range of spin magnitudes corresponding to bound violations is significantly broader for $L=5.00$ than for any other angular momentum value considered.

\begin{figure}[h]
	\centering
	\includegraphics[width=10cm,height=7cm]{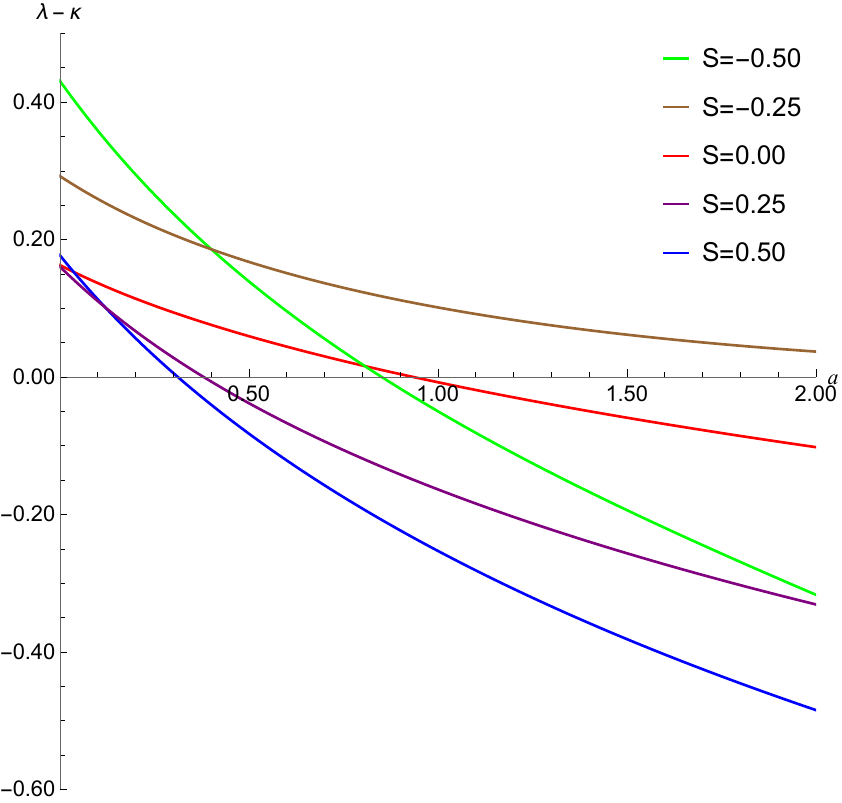}
	\caption{The relationship between the violation of the chaos bound and the EH constant. In the calculations, we set the parameters as follows: $L=10.00$, $\Lambda=-3.00$ and $q=-10.00$}
	\label{4f4}
\end{figure}

Figure \ref{4f4} demonstrates that the discrepancy between these exponents and the surface gravity decreases monotonically with increasing the EH constant. For the particles with identical spin orientations, the range of EH constant values corresponding to violations of the bound is broader when the spin magnitude is small compared to when it is large. In the case of $S = -0.25$, the figure indicates that the bound is consistently violated. Furthermore, the spin orientation also influences the threshold EH constant for the violation: the critical EH constant is smaller when the spin is parallel to the angular momentum than when it is antiparallel. Consequently, violations of the bound are more likely to occur when the spin orientation of the particle is opposite to the $z$-axis direction.

\begin{figure}[h]
	\centering
	\includegraphics[width=10cm,height=7cm]{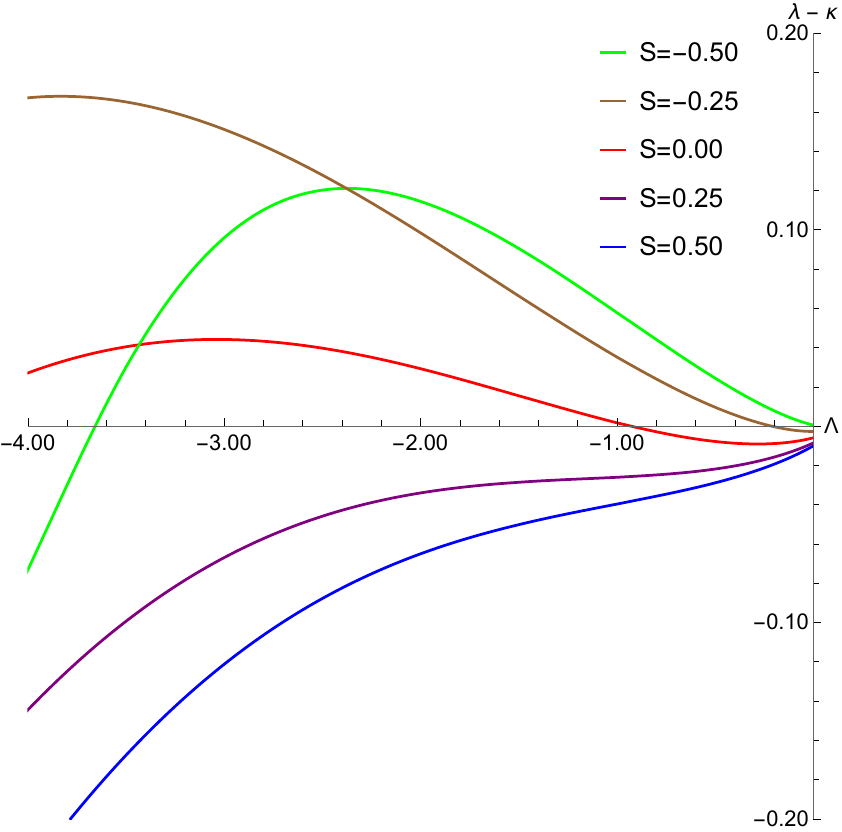}
	\caption{The relationship between the violation of the chaos bound and the cosmological constant. In the calculations, we set the parameters as follows: $L=10.00$, $q=-10.00$ and $a=0.60$.}
	\label{4f5}
\end{figure}

Since no unstable equilibrium orbits exist for the particle outside the event horizon when the cosmological constant $\Lambda$ is positive, Figure \ref{4f5} is limited to cases where $\Lambda \leq 0$. The figure reveals the following behavior. When the spin direction is parallel to the angular momentum, the differences between the exponent and the surface gravity grow as $|\Lambda|$ decreases. However, the chaos bound is never violated, regardless of the value of $\Lambda$. In contrast, violations do occur when the spin is anti-parallel to the angular momentum or when the particle is spinless. In these cases, the differences between the exponent and the surface gravity first increase and then decrease as $|\Lambda|$ increases. For the spinless particle, violations are observed only when $\Lambda$ lies below a certain threshold. For the spin $S=-0.50$, violations appear in the range $-3.18 < \Lambda \leq 0.00$. Given that the observed cosmological constant is an extremely small quantity, these violations effectively persist in the vicinity of both the EEH and EEH-AdS black holes.

\begin{figure}[h]
	\centering
	\includegraphics[width=10cm,height=7cm]{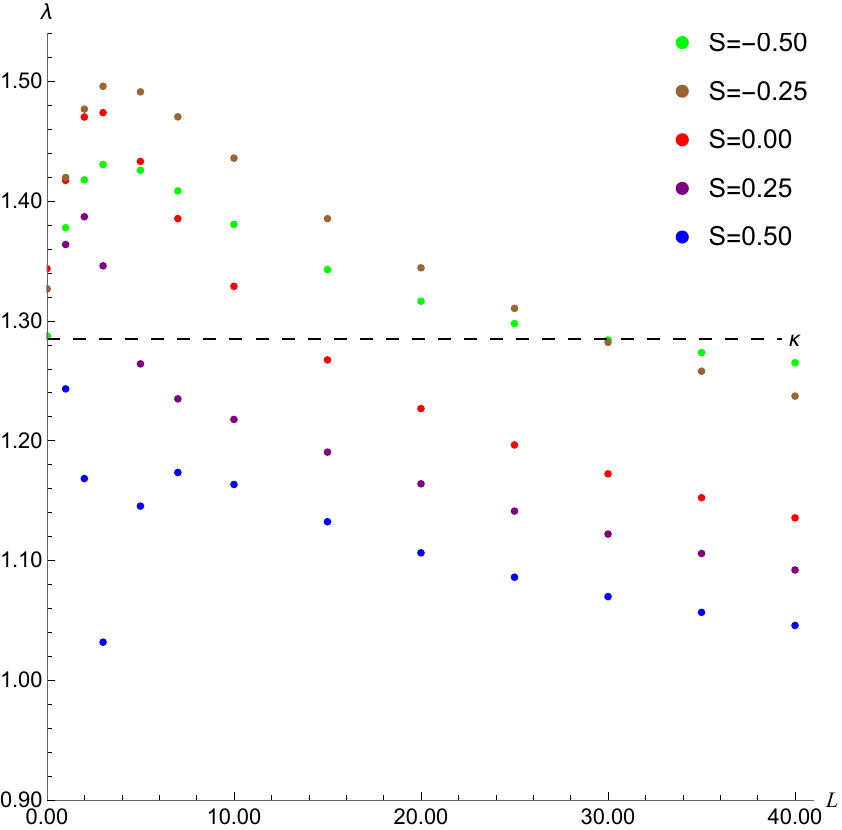}
	\caption{LEs vary with the particle's total angular momentum. In the calculations, we set the parameters as follows: $\Lambda=-3.00$, $q=-10.00$ and $a=0.60$.}
	\label{4f6}
\end{figure}

We further analyze the dependence of the exponents on the angular momentum of the particle in Fig. \ref{4f6}. The following observations can be made from the figure: Except for the case of the spin $S = 0.50$, the exponents first increase and then decrease with the increase of angular momentum. The violations occur when the angular momentum is less than the corresponding threshold. The range of angular momentum where the bound is violated gradually shrinks as the spin increases from the negative direction to the positive direction. This indicates that it is easier to induce a violation when the spin direction of the particle is opposite to the $z$-axis than when it is aligned with the $z$-axis. For the spin $S = 0.50$, the exponent exhibits two distinct extrema; one of these extrema exceeds the surface gravity, thus leading to a violation, while the other extremum does not exceed the surface gravity.

\begin{figure}[h]
	\centering
	\includegraphics[width=10cm,height=7cm]{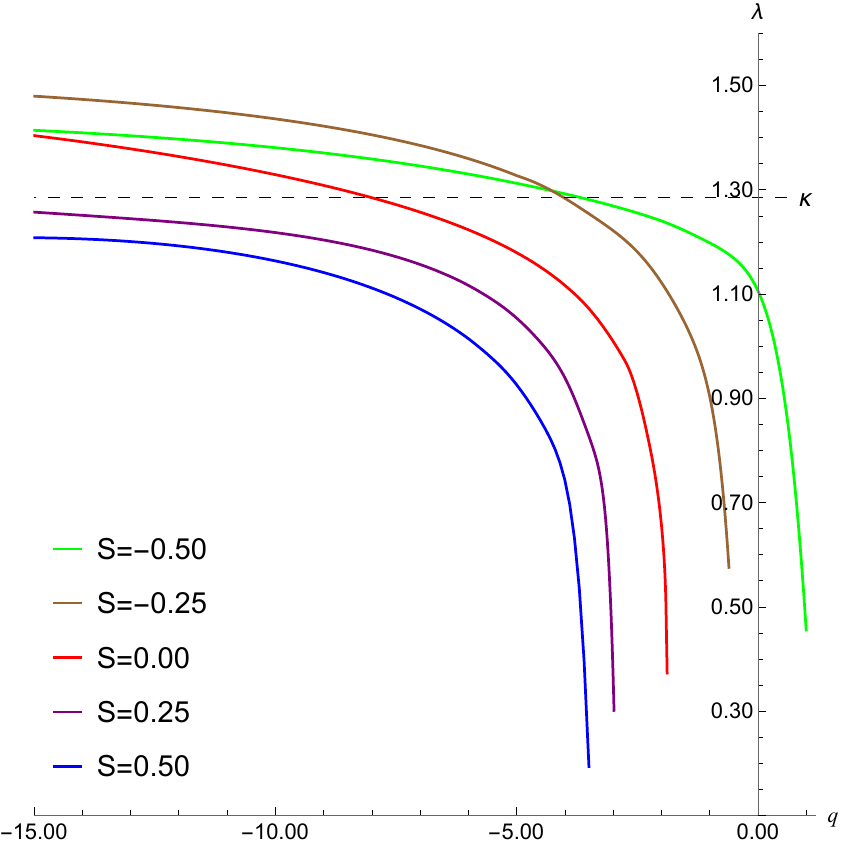}
	\caption{LEs vary with the particle's charge. In the calculations, we set the parameters as follows: $L=10.00$, $\Lambda=-3.00$ and $a=0.60$.}
	\label{4f7}
\end{figure}

Figure \ref{4f7} illustrates the dependence of the exponents on the particle's charge. As shown in the figure, the exponents increase monotonically with the magnitude of the negative charge. However, when the spin direction is parallel to the angular momentum direction, the exponents never exceed the surface gravity, and consequently no violation of the bound occurs. Violations are observed only when the spin is anti-parallel to the angular momentum or when the particle is spinless, and even in these cases only when the magnitude of the negative charge surpasses a certain threshold. For a positive charge, an exponent curve appears only for the anti-parallel spin configuration; in that case the exponent remains below the surface gravity. By contrast, when the spin is parallel to the angular momentum or when the particle is spinless, no stable equilibrium orbits exist outside the horizon, and therefore no exponents are defined.

\section{Conclusions}\label{sec4}

In this paper, we investigated the influence of key parameters-the black hole charge, EH constant, cosmological constant, the particle charge and its total particle angular momentum-on the LEs in the EEH-AdS spacetime. We found that violations of the chaos bound occur within the spinor field of this spacetime, driven by the combined effects of these parameters. The specific parametric dependencies are summarized as follows:

\begin{itemize}
	
	\item Dependence on the black hole charge, the particle spin and its total angular momentum: When other parameters are fixed-e.g., a specific particle spin or total angular momentum-the violations are observed only within the distinct ranges of black hole charge, particle spin, or total angular momentum. Notably, increasing the black hole charge or the particle's total angular momentum does not monotonically enhance the violation, a key distinction between the EEH-AdS spacetime and the RN spacetime \cite{Yang2026}.
		
	\item Dependence on the EH constant: With other parameters held constant, as the EH constant increases, the deviation between the exponents and the surface gravity gradually decreases, leading to a transition from chaos bound violation to compliance.
	
	\item Dependence on the cosmological constant and the spin orientation: At the fixed values of other parameters, the effect of the cosmological constant on the bound violation is sensitive to both the magnitude and direction of particle spin. When the spin is anti-aligned with the $z$-axis, even a small cosmological constant can trigger a violation. In contrast, when the spin is aligned with the $z$-axis, no violation occurs regardless of the cosmological constant's value.
	
   \item Role of the particle charge: The particle charge significantly influences the violations, indicating that the electromagnetic attraction can more readily trigger such violations than  the electromagnetic repulsion, provided the particle charge exceeds a critical threshold.
 \end{itemize}

It should be noted that the chaos bound was originally proposed in the context of thermal quantum systems, whereas our work explores the relationship between the LEs and surface gravity within a classical framework. Furthermore, when determining whether the bound is violated or not, genuine physical scenarios should be taken into account simultaneously, namely: The back-reaction of particle on the background spacetime needs to be considered \cite{LG1}. Meanwhile, the mass of the particles is much smaller than that of the black hole.

\vspace{10pt}

\noindent {\bf Acknowledgments}
This work is supported by the Sichuan Science and Technology Program (2024NSFSC1999).

\end{document}